\documentstyle[prl,aps,epsf,twocolumn]{revtex}

\begin{document}
\bibliographystyle{prsty}
\twocolumn[\hsize\textwidth\columnwidth\hsize\csname@twocolumnfalse\endcsname
%
\title{Self-Organized Criticality Driven by Deterministic Rules}
\author{Paolo De Los Rios, Angelo Valleriani and Jos\'e Luis 
Vega$^{\dagger}$}
\address{Max-Planck-Insitut f\"ur Physik Komplexer Systeme,
Bayreuther Str. 40, Haus 16, D-01187 Dresden.}
\maketitle

\begin{abstract}
We have investigated the essential ingredients allowing a system to show 
Self Organized Criticality (SOC) in its collective behavior. Using the 
Bak-Sneppen model of biological evolution as our paradigm,
we show that the random microscopic rules of update can be effectively
substituted 
with a chaotic map without changing the universality class. Using periodic
maps SOC 
is preserved, but in a  different universality class, as long as the spectrum
of frequencies is broad enough. 
\end{abstract}

\vspace{1cm}
\vfill
] 
\narrowtext 

Complex extended systems showing critical behavior, a lack of
scale in their features, appear to be widespread
in nature, being as diverse as earthquakes~\cite{CJ89}, creep
phenomena~\cite{Zaitsev92}, material fracturing~\cite{PPVAC94,DMP91,CTP96},
fluid displacement in porous media~\cite{WW83,CR88}, interface growth
~\cite{Sneppen92,Sneppen93}, river networks~\cite{RRRIB93,MCFCB96,CGMRR96} and 
biological evolution~\cite{Raup86,Gould77,EG88}.  
At variance with equilibrium statistical mechanics, these systems do
not need any fine tuning of a parameter to be in a critical state.
To explain this behavior, Bak, Tang and Wiesenfeld
introduced the concept of self-organized criticality (SOC) through
the simple sand-pile model~\cite{BTW87,BTW88}. 

In recent years, several models with extremal dynamics have been shown
to exhibit SOC when noise is present~\cite{BS93}. In this
 Letter we show that for this class of systems, noise can be replaced
by either a chaotic  or  a  {\it quasi-periodic} signal   
without destroying criticality. To illustrate this point we consider, as an
example, the model proposed by Bak and Sneppen (BS)  to describe the co-evolution
of natural species~\cite{BS93}. The result that different, even deterministic,
 microscopic rules can
induce SOC in the collective behavior of a population, points
to a greater relevance of SOC in nature.

In the BS model an ecosystem is described by a one-dimensional lattice, 
every site of
which is occupied by a species. Species with stronger mutual 
interactions in the ecosystem are arranged on nearest neighbor sites
(the lattice can be interpreted as a food-chain, or as a food-web
in more than one dimension). Each species is characterized by its fitness,
describing the average number of offsprings an individual of
that species can have in the given environment. This definition
of the fitness also accounts for the greater resistance to
mutations of fitter species since mutations must propagate over a greater 
number
of individuals to become a genetic trait of the species. Thus
the species with the lowest fitness is the one that feels the strongest
evolutionary pressure. Its fate is to either evolve or get extinct, 
and its place will be taken by some newcomer species in the same ecological 
niche.
Therefore the fitness of the species occupying that site is the
most likely to change in a short time. The nearest neighbor species will
find a different environment, and their fitnesses will result changed too.  As
a result of such a simple dynamical rule, the system exhibits 
sequences of causally connected evolutionary events called {\em
avalanches} \cite{BS93}. The number of avalanches $N$ follows a 
power law distribution
\begin{equation}
\label{ava}
N(s)\sim s^{-\tau}
\end{equation}
where $s$ is the size of the avalanche and $\tau\sim 1.07$ \cite{MPB94,PMB96} 
is the avalanche critical exponent.  
This kind of behavior,
which is the essence of self-organized criticality,
has actually been 
observed in paleontological data~\cite{Raup86} suggesting that
evolution and extinction may be episodic at all scales (a feature that goes
under the name of {\em punctuated equilibrium}) \cite{Gould77,EG88}.

In nature, the evolution of the least fit species is due to genetic mutation. 
In the BS model, this mutation is realized by giving to the corresponding species a random fitness.
As shown in \cite{BS93}, each lattice site $j$ is assigned a fitness,
namely a random number between $0$
and $1$.  At each time step in the simulation the smallest
fitness is found.  Then the fitnesses of the minimum and of the two nearest neighbors are 
updated  
according to the rule
\begin{equation}
f_{n+1}=F(f_{n}) \label{eq:noise}
\end{equation}
that assigns a new fitness $f_{n+1}$ at time $n+1$ to the chosen lattice site.
Indeed, in the original BS model, the function $F$ is just a random function with a
uniform distribution between 0 and 1. 
The system reaches a stationary critical state in which the distribution of
fitnesses is zero below a certain 
threshold $f_{c}\sim 0.66702$ \cite{MPB94,PMB96} and uniform above it. 
It is also possible to define other 
quantities that show a power law behavior with their own
critical exponent. 
Prominent among them are the first and all return time
distributions of activity (a site is defined as active when its fitness is
the minimum one),
\begin{equation}
\begin{array}{lcr}
\begin{displaystyle}
P_f(t)\sim t^{-\tau_f}\mbox{ }
\end{displaystyle}
& \mbox{ } &
\begin{displaystyle}\mbox{ }
P_a(t)\sim t^{-\tau_a}\, ,
\end{displaystyle}
\end{array}
\end{equation}
where $\tau_f\sim 1.58$ and $\tau_a\sim 0.42$ \cite{MPB94,PMB96}. 

Different microscopic rules have been proposed to describe how the mutation
of the least fit species induces mutations of the nearest neighbors
\cite{VDB96}: changing the microscopical dynamical rule affects the
universality class of the model, but not its SOC property. This fact shows
the robustness of the SOC behavior and poses the question of what are
the minimal requirements the BS model has to satisfy in order to be
critical. 

\begin{figure}[h]
\centerline{\epsfysize=7cm\epsfbox{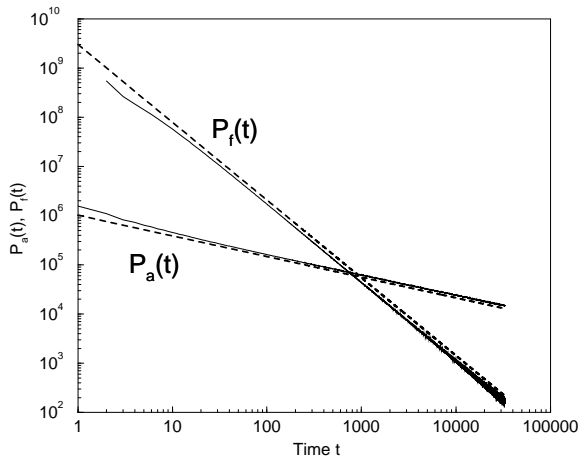}}
\caption{ First and all return distributions for a BS model with Bernoulli
updating rule with $r=2$.  
For all the simulations shown here, we used a lattice of $2^{14}$  sites and 
$5\times 10^{9}$ iterations exploiting the tree-algorithm explained in
[23]. 
\label{fig:1a}}
\end{figure}

\begin{figure}[h]
\centerline{\epsfysize=7cm\epsfbox{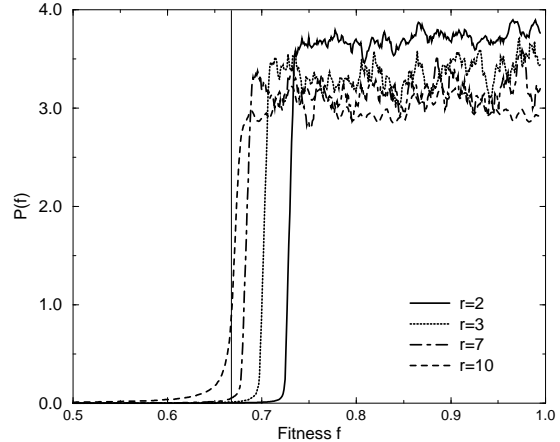}}
\caption{ Distribution of the fitnesses for $r=2,3,7,10$; the
threshold 
for $r=2$ is quite different from the usual BS threshold while the threshold
corresponding 
to $r=10$ is very close to the BS value (given by the vertical line).
For all the simulations shown here, we used a lattice of $2^{14}$  sites and $5\times 10^{9}$ iterations. 
\label{fig:1b}}
\end{figure}

While in the BS model the updating rule (\ref{eq:noise}) is a random
function, one could consider, instead,  deterministic updating. Indeed, we
started by considering a deterministic rule, whose statistical properties
resemble those of a stochastic function, namely  the Bernoulli map \cite{Schuster89}
\begin{equation}
\label{bernoulli}
f_{n+1}=G_{r}(f_{n})=[rf_{n}] \, ,
\end{equation}
where $[f]$ stands for the value of $f$ modulus $1$ and $r\in {\bf N}$ is a
constant. 
It has been shown (see \cite{Schuster89} and references therein) that this map
has a 
uniform invariant measure (for any integer value of $r$) and that the Lyapunov
exponent $\lambda$ 
is given by 
$\lambda= \log r$. 
In Fig.\ \ref{fig:1a} we show the power-law behavior of the first and all return
probability 
distributions in the case $r=2$.  The critical exponents  obtained coincide 
with those found in \cite{BS93} for the random updating. Moreover, our
simulations show that for
all values of $r$ the systems fall in the BS universality class, i.e. they all have 
the same critical exponents. 
The stationary distribution of the fitnesses,
on the other hand,
follows a different pattern. Indeed, Fig.\ \ref{fig:1b} shows that the  threshold for
$r=2$ is bigger 
than the one found for the random case.
On increasing the value of $r$, the threshold moves towards the BS value (see
Fig. \ref{fig:1b}). 
For non integer values of $r$ ($r>1$), SOC is still preserved
within the BS universality class. However, in this case, 
 the distribution of the generated numbers is not
uniform and consequently it  
influences the distribution of the fitnesses at the stationary state. 

\begin{figure}[h]
\centerline{\epsfysize=7cm\epsfbox{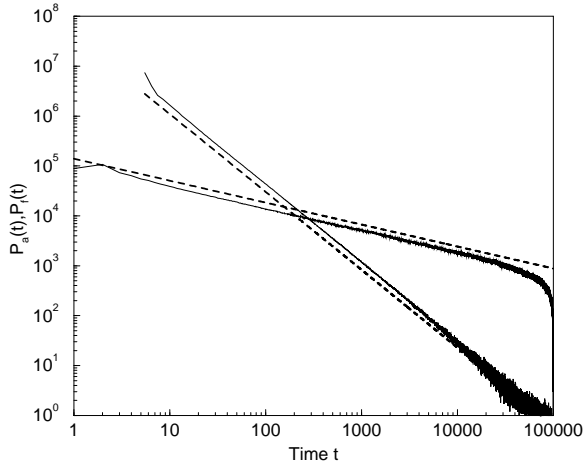}}
\caption{ First and all return distributions for a BS model with the
logistic map with $\lambda=4$ as updating rule. The exponents are the same as
for the BS model $\tau_f=1.67$ and $\tau_a=0.33$. In all the simulations shown
in this figure, we used a lattice 
of $2^{13}$ sites and $10^{8}$ iterations.  
\label{fig:2}}
\end{figure}

The  next step is then to consider updating rules that can be tuned 
to chaotic behavior by changing a  parameter.
To that effect, we take as updating rule for the
fitnesses the logistic 
or Feigenbaum 
map, namely 
\begin{equation}
\label{feig}
f_{n+1}= \lambda f_{n}(1-f_{n})\, .
\label{feigenbaum}
\end{equation}
The reasons for studying this rule are manifold. On the one hand, 
this map has already been considered in the context of biological
evolution models and population dynamics \cite{May74,May76,MO76} and can thus provide a 
possible deterministic interpretation
of the evolution inside every ecological niche. Moreover, it has been shown
that it describes the 
behavior of a wide variety of systems in nature \cite{CE80}. On the other hand, it has
a regime in which it is chaotic as well as one in which it is not,  
depending on whether $\lambda$ is bigger or less than the critical value
$\lambda_\infty\sim 3.56994$ \cite{Schuster89}. 
As Fig. \ref{fig:2} shows, for  those values of $\lambda$ for which the map is chaotic, 
 the system not only exhibits SOC but also stays in the same 
universality class as the original BS model. 
For $\lambda < \lambda_\infty$ we find that the system is not critical 
any more. Indeed below $\lambda_\infty$ every site follows a periodic orbit and 
to understand this loss of criticality 
one needs to investigate the case of periodic 
updating rules  in every site. 

Let us then consider a model in which the choice of the new fitness is
done according to the map
\begin{equation}
\begin{array}{c}
\begin{displaystyle}
f_{n+1}\, = \, \frac{sin(arcsin (2 f_{n}-1)+  \phi_{j} )+1}{2} \, = 
\end{displaystyle} \\ \\
\begin{displaystyle}
=\, \frac{sin(\omega_{j}  (n+1)
+ \phi_{j} )+1}{2}
\end{displaystyle}
\end{array}
\label{oscillators}
\end{equation}
where the $\omega_{j},\phi_{j}$'s are, in principle, 
different for each site $j$ and $n$ is the time step. 
In our calculations we have chosen  the frequencies $\omega_{j}$ such that
\begin{equation}
\omega_{j} \neq \omega_{i}
\label{disorder}
\end{equation} 
and the phases $\phi_{j}$ to verify
\begin{equation}
\phi_{j} \neq \phi_{i} .
\end{equation} 

\begin{figure}[h]
\centerline{\epsfysize=7cm\epsfbox{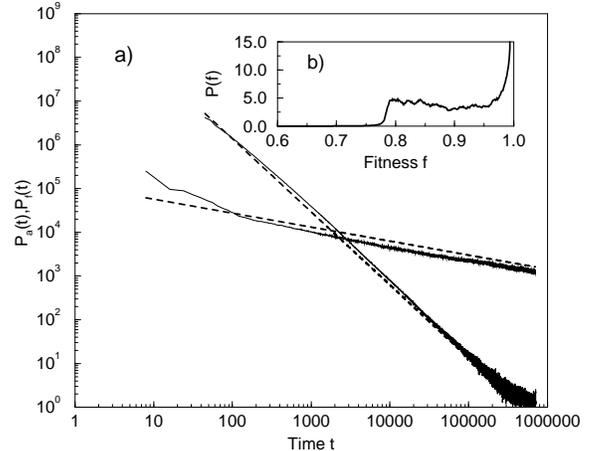}}
\caption{ (a) First and all return distributions for a BS model with disordered
periodic updating rules; the exponents are $\tau_f=1.67(1)$ and
$\tau_a=0.33(1)$. (b) Distribution of the fitnesses. In all the simulations
shown in this picture, we used a lattice of $2^{12}$ sites and $5\times 10^{9}$ iterations.  
\label{fig:3}}
\end{figure}

Bearing in mind that these maps are periodic, let us turn to Fig. \ref{fig:3} where 
the first and all return probabilities are shown. 
The universality class changes with respect to the original BS model,
with $\tau_f = 1.67(1)$ and $\tau_a = 0.33(1)$, but the SOC behavior is
preserved.
If instead of choosing all the
frequencies different we choose $\omega_{j}=\omega$ and all phases different
$\phi_{j} \neq \phi_{i}$ the SOC is destroyed (even if 
the fitnesses are organized above a threshold). 
This result sheds light on the loss of criticality for model
(\ref{feigenbaum})
when $\lambda < \lambda_\infty$, because there the system reduces to a set of 
oscillators, all with the same frequency and random initial phases.
Model (\ref{feigenbaum}) should recover the SOC behavior if
$\lambda$ is allowed to vary from site to site (mimicking condition 
(\ref{disorder}) of model
(\ref{oscillators})). 

It is worth noticing that even though chaotic maps
 produce  series of numbers that may (statistically) resemble random numbers
(with the exception of the functional form of the invariant density),
the behavior of the 
systems feels the details of the underlying dynamics, as shown by the 
dependence of the threshold on the parameter $r$ of the Bernoulli map. 
In particular,  the Bernoulli map for $r=2$ is 
formally equivalent to a coin toss \cite{Schuster89,Ford83} (the paradigm of 
randomness) 
and still  the threshold is different. 
However, at variance with the case of pure random noise, in the Bernoulli map
the correlations decay exponentially instead of 
being a delta function. 
Then the system feels the details of the
 rule and while staying in the same universality class exhibits a different threshold. 
One can then conclude that, provided the statistical properties are 
those of a random process, SOC will persist.   These results are
 complementary of the ones obtained 
in Ref.\ \cite{Bianucci93,Bianucci94}: They showed that 
a chaotic system can be used instead of a
heath bath to obtain thermalization (the chaotic system is referred to
as a ``booster'' \cite{Bianucci93,Bianucci94}). This is analogous
to what 
happens in the BS model: Noise (thermal or otherwise) 
can be replaced by a deterministic chaotic system without significant
 changes in the stationary state. However stochasticity in the updating rule
is sufficient but not necessary:   SOC persists even in the absence of chaos, for periodic updating rules.

Summarizing, the results shown here indicate that the feature that ensures SOC in systems with extremal dynamics, is not 
the randomness of the actual updated value  but the fact that the 
choice of the site  where the change is going to be performed 
(namely the minimum rule) is random. 
Moreover, as long as there is enough diversity among the species on the lattice,
the longer the memory (or the internal correlation) of each member, the higher the threshold. 
Indeed, in the case of chaotic maps, the diversity is ensured by the random
assignation of the initial 
values and as much as the chaoticity is increased we see that the threshold
decreases. The extremal case 
being given by the BS model.

In the case of the periodic map instead, the random initial conditions are not
enough to ensure diversity. 
Thus, in order to have SOC, we have to choose at random also the internal
time-scale {\it i.\ e.\ } the periods.
 
 These results add strength to the relevance of SOC in physics and biology,
since they allow different microscopic mechanisms to underlie
its appearance as a collective behavior.
From the point of view of biological evolution, this result could also account
for less wild variations 
of the fitnesses.
 
At this stage,  several questions arise.
What is the mechanism that allows the system to 
recognize the underlying dynamics? 
 Moreover, when the microscopic rule
is disordered periodic, does the distribution of the
frequencies influence the universality class of the model?

These questions are, we think, of utmost 
importance in order to understand the widespread appearance of self organized 
criticality in nature.\\

{\em Acknowledgments.}: We would like to thank T. Uzer, A. Maritan,
R. Cafiero and H. Kantz  for their 
useful comments and suggestions and S. Panzeri for a careful reading of the manuscript. 

Note: After finishing this work we became aware of reference \cite{CPD97}
where  a similar line of reasoning is pursued for the effects of disorder on a
population of integrate-and-fire oscillators.

\end{document}